\documentclass[12pt]{article}
\usepackage[dvips]{graphicx}
\def\thefootnote{\fnsymbol{footnote}}

\newcommand{\eq}{\begin{equation}}
\newcommand{\en}{\end{equation}}
\newcommand{\eqa}{\begin{eqnarray}}
\newcommand{\ena}{\end{eqnarray}}

\newcommand{\mc}{\multicolumn}

\begin{document}
\begin{titlepage}
\vskip0.5cm
\begin{flushright}
HUB-EP-99/07\\
\end{flushright}
\vskip0.5cm
\begin{center}
{\Large\bf A Monte Carlo study of leading order} 
\vskip0.3cm
{\Large\bf scaling corrections  of $\phi^4$ theory}
\vskip0.3cm
{\Large\bf on a three dimensional lattice}
\end{center}
\vskip 1.3cm
\centerline{
 M. Hasenbusch\footnote{e--mail: hasenbus@physik.hu-berlin.de}}
 \vskip 1.0cm
 \centerline{\sl Humboldt Universit\"at zu Berlin, Institut f\"ur Physik}
 \centerline{\sl Invalidenstr. 110, D-10115 Berlin, Germany}
 \vskip 1.cm

\begin{abstract}
We present a Monte Carlo study of the one-component
$\phi^4$ model on the cubic lattice in 
three dimensions. Leading order scaling corrections 
are studied  using the finite size scaling method.
We compute the corrections to scaling exponent $\omega$ 
with high precision. 
We determine the value of the coupling
$\lambda$ at which leading 
order corrections to scaling vanish. Using this result we  
obtain estimates for critical exponents  that are 
more precise than those obtained with field theoretic  
methods.
\vskip0.2cm
\end{abstract}
\end{titlepage}

\setcounter{footnote}{0}
\def\thefootnote{\arabic{footnote}}

\section{Introduction}
The divergence of quantities like the correlation length $\xi$ or 
the magnetic susceptibility $\chi$ in the neighbourhood of a 
critical point 
is described by scaling laws 
\begin{equation}
 \xi \propto t^{-\nu}   \;\;\;\;\;,  \;\;  \chi  \propto t^{-\gamma} ,
\end{equation}
where $t=|T-T_c|/T_c$ gives the distance from the critical point.
However, such scaling laws are valid in this simple form only in an
infinitesimal neighbourhood of the critical temperature on 
infinitely large systems.  
Monte Carlo simulations however are   
performed with finite systems. Therefore the analysis of the resulting 
data requires knowledge of corrections to scaling.  
A similar observation holds for experimental data of critical 
systems that are 
taken at a finite distance from the critical temperature.

While  renormalization group \cite{wegner}
(see also, e.g., ref. \cite{cardy})
predicts the structure 
of corrections qualitatively, an understanding on a quantitative level
is needed for the correct interpretation of Monte Carlo or experimental
data. From $\epsilon$-expansion,  perturbation theory in three 
dimensions and high temperature series expansions we know 
that leading corrections are proportional to
$\xi^{-\omega}$, with $\omega \approx 0.8$  for the universality class
of the three dimensional Ising model
\cite{WiFi,Parisi,guzi,ChFiNi,NiRe}.
The systematical error that is quoted for $\omega$ is about $1\%$ to $5\%$.
In a study \cite{scaling} of universal amplitude ratios 
of the Ising universality class
the uncertainty of the estimate of $\omega$ turned out to be
a major source of systematic errors.  
Recent Monte Carlo simulations \cite{spainparisi,us} 
indicate that the value of $\omega$ could
be considerably larger than $0.8$.
Little is known about sub-leading corrections 
to scaling.

Already in refs. \cite{ChFiNi,NiRe} it was suggested that the study 
of models that interpolate between the Gaussian model and the Ising model
should be used to study leading order corrections to scaling. 
Such models allow to vary the amplitude of the corrections to scaling.
In particular they allow to eliminate leading order corrections by 
a suitable choice of the parameters of the action (Hamiltonian).
This idea was recently implemented in the framework of Monte Carlo
simulations and finite size scaling \cite{us,spain}. 

In this paper we study the one-component $\phi^4$ (or Landau-Ginzburg)
model 
in three dimensions
on a simple cubic lattice. 
The action is given by
\eq
S = \sum_x \{- 2 \kappa \sum_{\mu} \phi_x \phi_{x+\hat \mu}  +\phi_x^2
 + \lambda (\phi_x^2 -1)^2 \} \;\; ,
 \label{action}
\en
where
the field variable $\phi_x$ is a real number and $x$ labels the lattice
sites. $\mu$ labels the directions and $\hat\mu$ is an unit-vector in 
$\mu$-direction. The Boltzmann factor is $\exp(-S)$.
For $\lambda=0$ we get the Gaussian model on the lattice. In the  
limit $\lambda=\infty$ the Ising model is recovered.
Following ref. \cite{spain} leading order scaling 
corrections vanish at $\lambda = 1.0(1)$.  
The authors of ref. \cite{us} find $\lambda\approx 1.145$.

The aim of the present study is three-fold:

We improve the accuracy of
the $\lambda$ at which leading 
order scaling corrections vanish. In particular, we give error estimates
for the value that is obtained.  

We obtain an accurate estimate of the correction 
exponent $\omega$. 
By simulating various values of $\lambda$ we are able to vary the 
strength of the leading order corrections. 

Finally simulations at the optimal $\lambda$ yield accurate 
results for the critical exponents $\nu$ and $\eta$. 

In section 2 we discuss how corrections to 
scaling that arise from the crossover
from the Gaussian fixed point to the Wilson-Fisher fixed point can be 
studied by
finite size scaling. The Monte Carlo algorithm is explained in section 3.
In section 4 we give an overview of the simulations that have been 
performed. The analysis of the data is presented in section 5.  
In section 6 we compare our  results  with the literature. Finally we 
give our conclusions and an outlook.

\section{Scaling corrections and finite size scaling}
$\epsilon$-expansion \cite{WiFi}
tells us that leading corrections to scaling are related 
to the RG-flow from the Gaussian fixed point 
into the Wilson-Fisher fixed point.
In Monte Carlo simulations of lattices with finite size $L$
the Binder cumulant 
\begin{equation}
 U(L,\kappa,\lambda) = \frac{<m^4>}{<m^2>^2}
\end{equation}
is the most natural quantity to monitor this flow. The magnetization
is given by  $m=\sum_x \phi_x$.

Since we like to study the flow on the critical surface it is useful
to consider 
a second phenomenological coupling (i.e.\ a non-trivial 
quantity that is invariant under RG-transformations).
We have chosen the ratio of partition 
functions with periodic and anti-periodic boundary conditions
$Z_a/Z_p$ \cite{mich}.  Note that partition functions are by construction
conserved under RG-transformations. Therefore also the 
ratio of partition
functions $Z_a/Z_p$ is conserved.

Instead of computing the Binder cumulant at 
$\kappa_c$, it is computed at the $\kappa$ for that $Z_a/Z_p$ takes a 
fixed value
on the given lattice. 
The practical advantage of this method is that
no errors are introduced by an inaccurate estimate of 
$\kappa_c$ and that due to cross-correlations
the statistical error of the Binder cumulant at fixed  $Z_a/Z_p$
is smaller than that of the Binder cumulant at fixed $\kappa$.
In the following we will always fix $Z_a/Z_p=0.5425$, which is according 
to ref. \cite{us} a good
approximation of 
\begin{equation}
\lim_{L\rightarrow \infty} Z_a/Z_p |_{\kappa_c} \;\; .
\end{equation}
The Binder cumulant at a fixed value of $Z_a/Z_p=0.5425$ means:
\begin{equation}
 \bar{U}(L,\lambda) = U(L,\bar{\kappa}(L,\lambda),\lambda) \;\; ,
\end{equation}
where $\bar{\kappa}(L,\lambda)$ is determined by
\begin{equation}
Z_a/Z_p(L,\bar{\kappa}(L,\lambda),\lambda)=0.5425 \; \; .
\end{equation}

The fact that there exists a unique RG-trajectory running
from the Gaussian fixed 
point into the Wilson-Fisher fixed point leads to
\begin{equation}
\label{sca}
\bar{U}(L,\lambda) =  f(a(\lambda) \; L) + b(\lambda) \;  L^{-x} + ... 
\;\; ,
\end{equation}
where we expect $x \approx 2$. The function $a(\lambda)$ goes to $0$ as 
$\lambda \rightarrow 0$. For some finite value of $\lambda$ leading order
scaling corrections vanish and 
$a$ diverges. Reparametrizing the scaling function $f$
\begin{equation}
 \tilde f(c(\lambda) \; L^{-\omega}) =  f(a(\lambda) \; L) \;\; 
\end{equation}
and Taylor-expanding yields
\begin{equation}
 \bar{U}(L,\lambda) = \bar{U}^* + c_1(\lambda) L^{-\omega} + 
		      c_2 \; c_1(\lambda)^2   L^{-2 \omega} + ... \;\; .
\end{equation}
In addition to the leading correction $L^{-\omega}$ and powers of it 
we should expect
that corrections of order $L^{-2}$, $L^{-4}$ and so on, which exist in
the lattice version of the Gaussian model, do survive in some 
form also at the Wilson-Fisher fixed point. One example of such a 
correction
is the restoration of the rotational invariance.
The result of ref. \cite{CaPeRoVi} indicates that such
corrections exist almost
unaltered at the Wilson-Fisher fixed point.

\section{The Monte Carlo algorithm}
We followed the idea of Brower and Tamayo \cite{BrTa}
and used a combination of the cluster algorithm and a local Metropolis
algorithm for updating the field. 
We replaced the Swendsen-Wang cluster
algorithm \cite{SW} that was used by Brower and Tamayo 
with the recently proposed wall-cluster algorithm \cite{us}.
The cluster algorithm only updates the sign of 
the field $\phi$.
Ergodicity of the update scheme is reached
by alternating the 
cluster update with Metropolis updates that allow to change the modulus 
of the field $\phi$. Below we give the details of the update schemes that
were used.

\subsection{Metropolis}
In order to make the updating scheme ergodic we performed one Metropolis
sweep in an update cycle. The proposal for the site $x$ is generated by
\begin{equation}
 \phi_x' = \phi_x +  s \left(r-\frac12 \right) \;\; ,
\end{equation}
where $r$ is a random number uniformly distributed in $(0,1]$ and 
$s$ parametrizes the size of the change. In our study we have chosen
$s=3$, yielding acceptance rates between $0.4$ and $0.6$, depending 
on $\lambda$.

\subsection{Overrelaxation}
In order to speed up the updating of the modulus of the field we 
added $n_o$ overrelaxation sweeps to the update cycle.
A proposal for the field at site $x$ is generated by
\begin{equation}
 \phi_x' = 2 \kappa \sum_{y.nn.x} \phi_y \;\; - \;\; \phi_x \;\; ,
\end{equation}
where $y.nn.x$ means that $y$ is a nearest neighbour of $x$.
This proposal keeps 
\eq
 S_{Gauss} =
 \sum_x \left[ - 2 \kappa \sum_{\mu} \phi_x \phi_{x+\hat \mu}  +\phi_x^2 
 \right]
\en
constant. 
This proposal is accepted if the demon variable  $d\in[0,\infty)$ can
take over the energy, i.e. $d' \ge 0$ with
\eq
 d' = d +  \lambda (\phi_x^2 -1)^2  - \lambda (\phi_x'^2 -1)^2 \;\; .
\en
This means that we keep the combined action 
$ S_{com} = S + d $
constant. $d$ is set equal to zero at the beginning of the simulation 
and it is then updated only with the overrelaxation updates.

As a check of the program we measured the expectation value of $d$:
\begin{equation}
< d > = \int_{0}^{\infty} \mbox{d}x \exp(-x) \; x  = 1 \;\; .
\end{equation}
For example from our simulation with $L=96$ at $\lambda=1.1$ we obtained
$< d > =1.000003(4) $.

The acceptance rate of the overrelaxation step depends on $\lambda$.  For 
$\lambda=0$ the acceptance rate is 1 by construction. For $\lambda=1.1$, 
where most of our simulations  are done, the acceptance rate 
is still $0.715$.

\subsection{Wall-cluster}
We followed the idea of Brower and Tamayo
and used the cluster-update only for updating the  sign of the 
field $\phi$. For that purpose
we consider the system as an Ising model with a link-dependent
coupling constant given by
$\beta_{<xy>} = 2 \kappa |\phi_x| |\phi_y| $.
This results in the link-dependent delete probability 
\eq
 p_d(x,y) = \mbox{min}[1, \exp(-2\beta_{<xy>} 
	   \; \mbox{sign}(\phi_x) \; \mbox{sign}(\phi_y))]
	  = \mbox{min}[1, \exp(-4\kappa \phi_x \phi_y)] \;\; .
\en
Given the delete probability there is still freedom in the 
choice of clusters to be flipped (i.e. the fields of the 
cluster are multiplied by $-1$). 
In the Swendsen-Wang \cite{SW} algorithm clusters are flipped with 
probability $1/2$. In Wolff's \cite{Wolff} single cluster algorithm 
the cluster that contains a randomly chosen 
site of the lattice is flipped. Other rules for the selection of 
the clusters to be flipped can be found in ref. 
\cite{Kerler}.
In this work we used the  wall-cluster algorithm introduced in 
ref. \cite{us}. In this case 
all clusters which intersect with a randomly chosen two-dimensional
plane of the 
lattice are flipped. In ref. \cite{us} we show that this version of 
the cluster algorithm is less affected by critical slowing down than 
the single cluster algorithm.

\section{The Simulations}
We performed simulations at a large range of $\lambda$ values and 
lattice sizes $L$. In table \ref{statist}
we give an overview of the simulation parameters
and the number of measurements for each set of simulation parameters.
Most of our simulations were performed on 200MHz Pentium Pro PCs 
running  under
Linux. The program is written in C. As random number generator 
we used our own implementation of G05CAF of the NAG-library.

\begin{table}[h]
\caption[blabla1]
{\label{statist}
\sl Summary of simulation parameters. In the first row we 
give the value of $\lambda$, in the second row the lattice size $L$ and 
in the third row the number of measurements divided by $3 \times 10^{6}$. 
}
\vskip 0.2cm
\begin{center}
\begin{tabular}{|l|l|l|}
\hline
$\lambda$ &  $L$                    & stat/$3  \times 10^{6} $   \\
\hline
0.1   & 6,8,12,16,24,32,48          & 6,5,5,5,4,2,1 \\ 
0.2   & 6,8,12,16,24,32,48          & 6,5,5,5,4,2,1   \\
0.4   & 6,8,10,12,14,16,18,20,24,32 & 6,5,5,6,9,6,6,5,5,2 \\
0.7   & 2,3,4,5,6,8,12,16,24        & 6,6,6,6,6,6,7,6,7 \\
0.8   & 2,3,4,5,6,8,12,16,24        & 6,6,6,6,6,6,5,5,5 \\
0.9   & 2,3,4,5,6,7,8,9,10,12,14,16 & 10,10,10,10,30,30,20,20,20,20,20,20\\
1.1   & 2,3,4,5,6,7,8,9,10,11,12,13,14,15,& 
	     8,8,8,8,25,30,30,32,30,11,10,6,6,6,\\
      & 16,18,20,22,24,28,32,40,48,64,96 &
        5,5,5,5.5,4,6,4,2.5,2.6,1,0.5 \\
1.145 & 2,3,4,5,6,7,8,9,10,11,12,13,14,& 
	8,8,8,8,25,30,30,30,30,11,10,6,6,\\ 
      & 15,16,18,20,24,32,48,64    & 6,5,5,6,4,3,1.7,0.5 \\
1.3   & 2,3,4,5,6,7,8,9,10,12,14,16&10,10,10,10,30,30,20,20,20,20,20,20\\
1.4   & 2,3,4,5,6,7,8,12,16,24     & 6,6,6,6,6,6,6,7,6,6 \\
1.5   & 2,3,4,5,6,7,8,12,16,24     & 6,6,6,6,6,6,5,5,5,5 \\
2.5   & 6,8,12,16,24,32            & 6,5,5,5,4,2 \\
\hline
\end{tabular}
\end{center}
\end{table}

Per measurement we performed one Metropolis sweep, one or
two overrelaxation sweeps and four or seven wall-cluster updates. The last
wall-cluster update in this sequence is also used 
for the measurement of the
boundary variable. The direction of the plane used for the wall-cluster 
is taken from a fixed sequence. First perpendicular to the 1-direction,
then perpendicular to the 2-direction, and then perpendicular to the 
3-direction and so on. The position of the plane is chosen randomly.

The update of a single site with Metropolis takes $2.0 \times 10^{-6}
sec$ and the overrelaxation $0.4 \times 10^{-6} sec$. 
The construction of one site of the cluster takes on average
$4.5  \times 10^{-6} sec$.

We computed the integrated
autocorrelation times of the observables that were measured.
For example for our largest lattice size $L=96$ at $\lambda=1.1$ and 
$\kappa \approx \kappa_c$,
using one Metropolis sweep, two
overrelaxation sweeps and
seven wall-cluster updates per measurement,
we obtain $\tau_{\chi}=3.56(2)$
and $\tau_b=1.93(1)$ in units of measurements for the integrated
autocorrelation time of the magnetic susceptibility $\chi$ and the
boundary variable, respectively.

The average sum of the sizes of the clusters per volume that are
flipped in one update step are fitted with the ansatz $S/V = C L^x$.
We obtain  $S/V = 1.394(4) \; L^{-0.4874(7)}$ 
taking into account $L=48$, $64$
and $L=96$ for $\lambda=1.1$ and $\kappa \approx \kappa_c$. The
$\chi^2$/d.o.f. = 4.9 is rather large. Taking only $L=64$ and $L=96$ 
yields $S/V = 1.410(8) \; L^{-0.4901(14)}$.

This result has to be compared with $S/V = 1.008(4) \; L^{-0.527(1)}$ that
was found in ref. \cite{us} for the standard Ising model on the cubical
lattice.
It seems that in the case of the $\phi^4$ model there are rather strong
corrections to the simple power law.

The simulations were performed at $\kappa_s$ 
that were the best estimates of 
$\bar{\kappa}$ 
available at the start of the simulations. In order to evaluate 
observables at $\kappa$ values different from $\kappa_s$ we used second 
order Taylor expansion in $\kappa$ at $\kappa_s$. We always checked that
the error made by truncating the Taylor series is well below the 
statistical error. The Taylor coefficients are obtained from the 
simulations.

In total we used about 3.5 years of Pentium Pro CPU-time for these 
simulations. The runs for 
$\lambda=1.1$, which turned out to be close to the optimal $\lambda$,
took about 570 days.

\section{Analysing the data}
\subsection{The Binder cumulant and corrections to scaling}
We analysed the Binder cumulant at $Z_a/Z_p=0.5425$ fixed in 
order to study corrections to scaling and to find the value of $\lambda$
where leading order corrections to scaling vanish. To get a first
impression we have plotted our data for $\lambda=0.4, 0.8, 1.1, 1.5$ and 
$2.5$ for $ 5 \le  L \le 24$ in figure 1. For $\lambda=0.4$ and 
$\lambda=0.8$, $\bar{U}$ is decreasing with increasing lattice size $L$. 
For $\lambda=1.1$ starting from about $L=10$ the value of $\bar{U}$ stays
constant within error-bars. Therefore the value of $\lambda$ at which
leading order corrections to scaling vanish should be very close
to $1.1$.
Going to  $\lambda=1.5$ and 
$\lambda=2.5$ the value of $\bar{U}$ is increasing with the lattice 
size. For a given lattice size $\bar{U}$  is monotonically increasing
with decreasing $\lambda$. 

\begin{figure}
\begin{center}
\includegraphics[width=12cm]{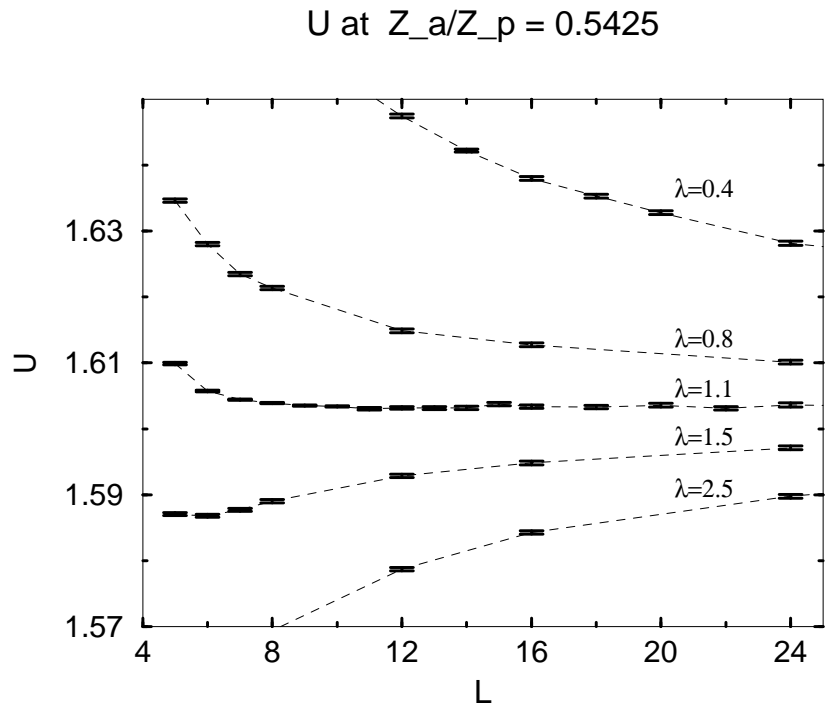}
\parbox[t]{.85\textwidth}
 {
\caption[Binder Cumulant $U$ at $Z_a/Z_p$ fixed 1]
 {\label{binderfig1} \small
 The Binder cumulant $U$ at $Z_a/Z_p=0.5425$ as a function of 
 the lattice size $L$ for $\lambda=0.4, 0.8, 1.1, 1.5$ and $2.5$.
}}
\end{center}
\end{figure}
Next we analyzed our data for $\bar{U}$ in a more quantitative fashion.
In a first attempt we fitted the data with the simple ansatz
\begin{equation}
\label{simple}
  \bar{U}(L,\lambda) = \bar{U}^* + c_1(\lambda) \;\; L^{-\omega} \;\; .
\end{equation}
Our fit results are summarized 
in table \ref{omega1}. The data which are included into the fit are 
determined by two criteria. First the lattice size has to be larger
than a minimal lattice size $L \ge L_{min}$.
Second the  distance $d=|\bar{U}(L,\lambda)-\bar{U}^*|$ has to be smaller
than a maximal distance $d \le d_{max}$. 
The first criterion allows to control effects of higher order
corrections in general, while the second specifically controls higher 
corrections that involve the same scaling field as the leading correction
($L^{-2 \omega}$, $L^{-3 \omega}$,  ...).

\begin{figure}
\begin{center}
\includegraphics[width=12cm]{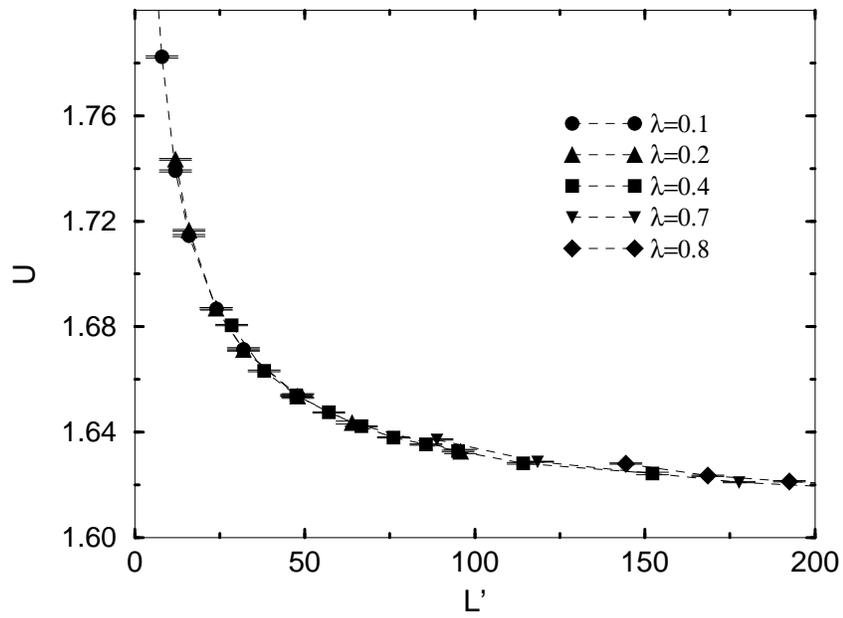}
\parbox[t]{.85\textwidth}
{
\caption[Binder Cumulant $U$ at $Z_a/Z_p$ fixed 2]
{\label{binderfig2} \small
 The Binder cumulant $U$ at $Z_a/Z_p=0.5425$ as a function of 
 the rescaled lattice size $L'=a(\lambda) L$ 
 for $\lambda=0.1, 0.2, 0.4, 0.7$ and $0.8$.
}}
\end{center}
\end{figure}

The numbers for $\chi^2/$d.o.f. indicate that at least $L_{min} = 12$ and 
$d_{max}=0.04$ is required to obtain a reliable fit. 
Results from the fits with a $\chi^2/$d.o.f. close to one
give $\omega=0.85$  up to $0.87$ for the correction to scaling exponent.
This value is clearly larger than that given by most recent field 
theoretic work \cite{guzi}.

Values for $c_1(\lambda)$ are not listed here. They will be discussed 
in more detail below.

In a second attempt  
the data were fitted with the ansatz
\begin{equation}
\label{simple2}
 \bar{U}(L,\lambda) = \bar{U}^* + c_1(\lambda) \;\; L^{-\omega}
		     + c_2 \; c_1(\lambda)^2  \;\; L^{-2 \omega} \;\;.
\end{equation}
The additional term requires only one further parameter 
in the fit. Our results are summarized in table \ref{omega2}.
We note that,
as expected, data with much larger values of $d=\bar{U}-\bar{U}^*$ 
can be included 
in the fit. The variation of $\omega$ with varying   
$L_{min}$ and $d_{max}$ is considerably reduced compared with the 
previous fit.

As our final result we quote $\omega=0.845(10)$, where we take into
account the variation of the result with $L_{min}$ and $d_{max}$.

For a small number of fits we give in table \ref{constant}
the results for $c_1(\lambda)$.
We see that for $\lambda=1.1$, within error-bars, the leading corrections
to scaling vanish. Converting the error-bar of $c_1(\lambda)$ to 
$\lambda$ we get $\lambda_{opt} = 1.100(7)$ from $L_{min}=12$, 
                 $\lambda_{opt} = 1.102(8)$ from $L_{min}=14$, and
                 $\lambda_{opt} = 1.095(12)$ from $L_{min}=16$.
		 $\lambda_{opt}$ is the $\lambda$ where leading order 
		 corrections vanish.

\begin{table}[h]
\caption[blabla2]
{\label{omega1} \sl
 Fit results for the Binder cumulant evaluated at $Z_a/Z_p=0.5425$ 
 fixed. The ansatz is given in eq. (\ref{simple}). 
 We  give results for various minimal lattice sizes $L_{min}$ 
 and maximal distances $d_{max}$ of $\bar{U}$ from its fixed point 
 value $\bar{U}^*$. $\omega$ is the correction to scaling exponent.}
\vskip 0.2cm
\begin{center}
\begin{tabular}{|c|c|c|c|c|}
\hline
$L_{min}$ &$d_{max}$ & $\chi^2$/d.o.f. &  $\bar{U}^*$  &  $\omega$ \\
\hline
\phantom{1}8&0.06 &5.36 & 1.60351(8)\phantom{1} & 0.865(3)\phantom{1}\\
\phantom{1}8&0.04&2.45  & 1.60300(8)\phantom{1} & 0.866(5)\phantom{1}\\
\phantom{1}8&0.03&2.06  & 1.60286(8)\phantom{1} & 0.866(7)\phantom{1}\\ 
 12    & 0.06   & 2.02  &  1.60398(13) & 0.840(5)\phantom{1} \\
 12    & 0.04   & 1.43  &  1.60370(14) & 0.856(6)\phantom{1} \\
 12    & 0.03   & 1.21  &  1.60344(15) & 0.864(10) \\
 16    & 0.06   & 1.85  &  1.60435(20) & 0.817(8)\phantom{1} \\   
 16    & 0.04   & 1.31  &  1.60379(22) & 0.853(11) \\
 16    & 0.03   & 1.18  &  1.60362(23) & 0.863(12) \\
\hline
\end{tabular}
\end{center}
\end{table}

\begin{table}[ht]
\caption[blabla3]
{
\label{omega2} \sl
 Fit results for the Binder cumulant at $Z_a/Z_p=0.5425$ fixed. 
 Compared with the previous table a term proportional to $L^{-2 \omega}$
 has been included into the ansatz (\ref{simple2}).
 }
\vskip 0.2cm
\begin{center}
\begin{tabular}{|c|c|c|c|c|c|}
\hline
$L_{min}$ & $d_{max}$  & $\chi^2$/d.o.f. &  $\bar{U}^*$ & $\omega$ & $c_2$ \\
\hline
  \phantom{0}8    & all    & 3.08 & 1.60318(8)\phantom{1} & 
   0.847(2)\phantom{1} &--1.01(2)\phantom{1}  \\
  \phantom{0}8    & 0.07   & 1.83 & 1.60276(9)\phantom{1} & 
  0.847(3)\phantom{1} & --1.64(7)\phantom{1}  \\
  \phantom{0}8    & 0.05   & 1.94 & 1.60277(9)\phantom{1} & 
  0.849(4)\phantom{1}  & --1.57(15) \\
 12    & all    & 1.19 & 1.60355(13)& 0.839(5)\phantom{1} &
 --1.17(5)\phantom{1}  \\
 12    & 0.07   & 0.95 & 1.60328(15)& 0.846(6)\phantom{1} & --1.57(10)  \\
 12    & 0.05   & 1.01 & 1.60328(16)& 0.846(6)\phantom{1} & --1.54(29)  \\
 16    & all    & 1.02 & 1.60363(21)& 0.839(9)\phantom{1} & --1.31(10)  \\
 16    & 0.07   & 0.95 & 1.60344(23)& 0.845(11)& --1.65(18)  \\
 16    & 0.05   & 1.02 & 1.60336(26)& 0.845(11)& --1.97(48)  \\
\hline
\end{tabular}
\end{center}
\end{table}

\begin{table}[h]
\caption[blablax]
{
\label{constant} \sl
 The correction to scaling amplitude $c_1(\lambda)$ as a function
 of $\lambda$ from fits with the ansatz (\ref{simple2}).
 We give the results for
 three values of $L_{min}=12,14$ and $16$. $d_{max}=0.07$ in all three
 cases.
 } 
\vskip 0.2cm
\begin{center}
\begin{tabular}{|c|c|c|c|c|c|}
\hline
 $\lambda$& $L_{min}=12$ & $L_{min}=14$ &
 $L_{min}=16$ \\
\hline
   0.1           &\phantom{---}1.452(44)\phantom{12}&
	\phantom{---}1.454(61)\phantom{12} & 
	\phantom{---}1.465(81)\phantom{12} \\
   0.2           &\phantom{---}0.809(21)\phantom{12} &
	\phantom{---}0.810(28)\phantom{12} & 
	\phantom{---}0.812(37)\phantom{12} \\
   0.4           &\phantom{--}0.3882(85) & 
	  \phantom{---}0.387(12)\phantom{12} 
	& \phantom{---}0.387(16)\phantom{12} \\
   0.7           &\phantom{--}0.1485(39) & \phantom{--}0.1482(52) &
		  \phantom{--}0.1462(72)\\
   0.8           &\phantom{--}0.0985(32) & \phantom{--}0.0982(42) &
		  \phantom{--}0.0988(59)\\
   0.9           &\phantom{--}0.0628(22) & \phantom{--}0.0623(31) & 
		  \phantom{--}0.0602(41)\\
   1.1           &\phantom{--}0.0001(18) & \phantom{--}0.0006(25) &
	       --0.0015(34)\\
\phantom{12}1.145&--0.0116(16) &--0.0128(22) &--0.0139(31)\\
   1.3           &--0.0479(13) &--0.0481(16) &--0.0482(23)\\
   1.4           &--0.0684(18) &--0.0686(19) &--0.0739(30)\\
   1.5           &--0.0857(20) &--0.0859(22) &--0.0896(32)\\
   2.5           &--0.1933(28) &--0.1934(36) &--0.1944(49)\\
 $\infty$        &--0.3112(39) &--0.3113(58) &--0.3125(76)\\
\hline
\end{tabular}
\end{center}
\end{table}

In order to check the existence of the scaling function of eq. (\ref{sca})
we have plotted in figure 2 $\bar{U}$ for $\lambda=0.1,0.2,0.4,0.7$, and 
$0.8$ as a function of 
\begin{equation}
L'= \left(\frac{c(\lambda)}{c(0.1)}\right)^{-1/\omega} \;\;  L \;\; .
\end{equation}
The values for $c$ are taken from table \ref{constant}. We see that 
almost all
data-points fall nicely on a unique curve. Only for the smallest lattice 
sizes that have been included ($L=6$) a small deviation is visible.

Finally we studied differences of the Binder cumulant at $Z_a/Z_p=0.5425$ 
fixed for different values of $\lambda$. We define
\begin{equation}
\Delta \bar{U} (L,\lambda_1,\lambda_2) =
\bar{U} (L,\lambda_1) - \bar{U} (L,\lambda_2) \;\; .
\end{equation}
This way $\bar{U}^*$ is cancelled and we expect
\begin{equation}
\label{diff}
\Delta \bar{U} (L,\lambda_1,\lambda_2) =
[c_1(\lambda_1)-c_1(\lambda_2)] \;\; L^{-\omega} + ... \;\;.
\end{equation}
As an example we give in table \ref{omega3} the results of fits  for 
$\lambda_1=0.9$ and $\lambda_2=1.3$. The interesting fact is that
already starting from $L_{min}=4$ the data are well fitted by the simple 
ansatz. Also the value for $\omega$ obtained this way is compatible 
with the result obtained above.
This means that sub-leading corrections  
depend very little on  
$\lambda$ and are, to a large extend, cancelled in $\Delta \bar{U}$.
Hence there should be a good chance to study these
corrections using perturbation theory.

Here we try to obtain a better understanding of the sub-leading 
corrections by 
looking at the results for small $L$ at 
$\lambda=1.1$ in more detail. Corrections to scaling vanish very rapidly.
Starting from $L=9$ the result for the Binder cumulant at $Z_a/Z_p$ fixed 
is consistent with $\bar{U}^*$ within the error-bars.
Fitting 
\begin{equation}
 \bar{U}(L) - \bar{U}^* = c \;\; L^{-x} \;\; ,
\end{equation}
we find $x\approx 6 \pm 1$  when including lattices of size $L=4$ up to 
$L=8$.  It is quite surprising that the numerically dominant 
corrections are governed by such a large exponent and not, as one might 
expect, by $x\approx 2$. 

Also the authors of  ref. \cite{blh} found that a correction term 
with a rather large exponent has to be included into the fit ansatz
in order to fit the Binder cumulant on small lattices.  They suggest 
that this correction exponent is given by $2 y_h$, 
where $y_h \approx 2.48$ 
is the RG-exponent
related with the external field.  To check this hypothesis we also looked 
at the corrections of $Z_a/Z_p$ at $\kappa_c$. It turns out that they 
vanish as rapidly as the corrections of the Binder cumulant.  Hence the 
observed corrections are related to an irrelevant  RG-exponent.

\begin{table}[h]
\caption[blabla4]
{
\label{omega3} \sl
 Fits of the difference
 of the Binder cumulant at $\lambda_1=0.9$ and $\lambda_2=1.3$ with the 
 ansatz (\ref{diff}). We give 
 $dc=c_1(\lambda_1)-c_1(\lambda_2)$ and the correction to scaling 
 exponent $\omega$. }
\vskip 0.2cm
\begin{center}
\begin{tabular}{|c|c|c|c|}
\hline
$L_{min}$  &   $\chi^2$/d.o.f. &  d$c$ & $\omega $  \\
\hline
   2     &     2.84       &  0.1149(7)\phantom{1} & 
				    0.870(4)\phantom{1}  \\
   3     &     2.83       &  0.1132(12)& 0.862(6)\phantom{1}  \\
   4     &     1.18       &  0.1078(18)& 0.838(8)\phantom{1}  \\
   5     &     1.26       &  0.1065(23)& 0.833(11)\\
   6     &     1.39       &  0.1076(28)& 0.837(12)\\
\hline 
\end{tabular}
\end{center}
\end{table}

\subsection{The critical line $\kappa_c(\lambda)$}

In order to obtain a result for the critical coupling $\kappa_c$ 
we fitted our data for the ratio of partition functions $Z_a/Z_p$ 
with the ansatz
\begin{equation}
 Z_a/Z_p(L,\kappa,\lambda) = Z_a/Z_p^* + 
 \frac{\partial Z_a/Z_p}{\partial \kappa}
  \; (\kappa-\kappa_c)  \;\; .
\end{equation}
For $\lambda=1.1$ we obtain with $L_{min}=12$  the result
$Z_a/Z_p^* = 0.54243(9)$  and $2 \kappa_c = 0.3750965(4)$.  
Increasing the minimal lattice size that is included in the fit
to $L_{min}=16$ we obtain 
$Z_a/Z_p^* = 0.54244(14)$  and $2 \kappa_c = 0.3750966(4)$.
In both cases  $\chi^2/$d.o.f. is a little larger than $1/2$.

The result  $Z_a/Z_p^* = 0.54244(14)$ is consistent with 
previous estimates 
$Z_a/Z_p^* = 0.54334(26)\{27\}$  from the standard Ising model and 
$Z_a/Z_p^* = 0.54254(14)[6]\{14\}$ from the spin-1 Ising model \cite{us}.
The numbers given in the second and third bracket are estimates of 
systematic errors caused by subleading corrections. In this work we 
skipped a detailed analysis of systematic errors. However it is reasonable
to assume that systematic errors are of similar size as for the spin-1
Ising model.
The estimate  $2  \kappa_c=0.3750966(4)$ for $\lambda=1.1$ will be useful 
in forthcoming studies of scaling laws.

For the other values of $\lambda$ we determined $\kappa_c$ from
\begin{equation}
Z_a/Z_p + \frac{\partial Z_a/Z_p}{\partial \kappa} 
\; (\kappa_c-\kappa) = 0.5425 \;\; .
\end{equation}
The results are \\
    $[0.1, 0.3734095(13)]$,
    $[0.2, 0.3884251(14)]$, 
    $[0.4, 0.3975837(13)]$,
    $[0.7, 0.3925302(20)]$,  
    $[0.8, 0.3887757(21)]$,
    $[0.9, 0.3845113(36)]$,
    $[1.145, 0.3728926(7)]$,  
    $[1.3, 0.3652233(33)]$,
    $[1.4, 0.3602789(19)]$,
    $[1.5, 0.3553854(21)]$,
    and 
    $[2.5, 0.3134347(16)]$
for the pairs $[\lambda, 2 \kappa_c]$




\subsection{The magnetic susceptibility}
As definition of the magnetic susceptibility we used
\begin{equation}
\chi = \frac{1}{L^3} \left( \sum_x \phi_x \right)^2  \;\; .
\end{equation}
Following refs.\ \cite{us,Ball} 
we tried to extract the exponent $\eta$ from the magnetic susceptibility
evaluated at  $Z_a/Z_p=0.5425$ or at $U = 1.6034$ fixed. In the following 
the magnetic susceptibility evaluated at  $Z_a/Z_p$ 
or at $U$ fixed is denoted by $\bar{\chi}$.
We fitted our data with the ansatz
\begin{equation}
\label{chifit1}
\bar{\chi}~(L) =  c +  d  \,  L^{2-\eta} \;\; , 
\end{equation}
where $c$ is an analytic correction.  Note that also corrections  that
decay like $L^{-x}$ with $x\approx 2$ 
are effectively parametrized by this ansatz.
\begin{table}[h]
\caption[blabla5]
{
\label{eta1} \sl
 Results for $\eta$ from  fitting the magnetic susceptibility at 
 $Z_a/Z_p = 0.5425$ fixed with the ansatz (\ref{chifit1}).}
  \vskip 0.2cm
  \begin{center}
   \begin{tabular}{|c|c|c|c|c|}
   \hline
 $L_{min}$ & $\chi^2/$d.o.f. & $\eta$  &  $d$   &  $c$ \\
   \hline
\phantom{0}4 & 25.05\phantom{0} & 0.0392(1) & 0.9856(3)\phantom{0} 
  & --0.724(4)\\
\phantom{0}6 &  2.73    &  0.0371(1) & 0.9789(4)\phantom{0} &
--0.577(8) \\
\phantom{0}8 &  1.23    &  0.0364(2) & 0.9765(7)\phantom{0}
&  \phantom{0}--0.504(18)\\
  10    &  1.23    &  0.0361(3) & 0.9755(9)\phantom{0} 
   &  --0.46(3)\phantom{0} \\
  12    &  1.33    &  0.0358(4) & 0.9745(14)&  --0.40(7)\phantom{0} \\
  14    &  1.29    &  0.0357(5) & 0.9740(18)&  --0.35(11) \\
  16    &  1.00    &  0.0358(6) & 0.9742(21)&  --0.35(17) \\
\hline
\end{tabular}
\end{center}
\end{table}

\begin{table}[h]
\caption[blabla6]
{
\label{eta2} \sl
 Results for $\eta$ from  fitting the magnetic susceptibility at 
 $U = 1.6034$ fixed with the ansatz (\ref{chifit1}).}
\vskip 0.2cm
\begin{center}
\begin{tabular}{|c|c|c|c|c|}
\hline
$L_{min}$ & $\chi^2/$d.o.f. & $\eta$  &  $d$   &  $c$ \\
\hline
 \phantom{0}4    &  3.24     & 0.0345(2)  &  0.9701(5)\phantom{0} & 
  --0.31(1)\phantom{0} \\
 \phantom{0}6    &  1.24     & 0.0356(3)  &  0.9736(7)\phantom{0} &  
 --0.38(1)\phantom{0}  \\
 \phantom{0}8 &1.34 & 0.0359(4) & 0.9744(12) & --0.41(3)\phantom{0} \\
10    &  1.28     & 0.0365(5)   & 0.9767(17) & --0.50(6)\phantom{0} \\
12    &  1.09     & 0.0370(7)   & 0.9784(25) & --0.59(12) \\
14    &  1.02     & 0.0368(9)   & 0.9768(32) & --0.45(20) \\
16    &  1.26     & \phantom{0}0.0366(10) & 0.9768(39) & --0.45(31) \\
\hline
\end{tabular}
\end{center}
\end{table}

First we performed fits for $\lambda=1.1$, where leading corrections
to scaling vanish. Results for  various $L_{min}$  are given in table
\ref{eta1} for $Z_a/Z_p=0.5425$ fixed
and table \ref{eta2} for $U=1.6034$ fixed. As before, $L_{min}$ gives 
the smallest lattice size that has been taken into account for the fit.
The $\chi^2/$d.o.f. becomes about $1$ for $L_{min}=8$  for
$Z_a/Z_p$ fixed, and  for $L_{min} =6 $ for $U$ fixed.

As a check we performed fits without analytic part $c$ for 
$Z_a/Z_p=0.5425$ fixed 
\begin{equation}
\label{chifit2}
\bar{\chi}~(L) =  d  \,  L^{2-\eta} \, .
\end{equation}
The results are summarized in table \ref{eta3}.  Starting from 
$L_{min}=28$ the result for $\eta$ becomes consistent with the results
obtained from the fit   eq. (\ref{chifit1})
that includes analytic corrections.

\begin{table}[h]
\caption[blabla7]
{
\label{eta3} \sl
 Results for $\eta$ from  fitting the magnetic susceptibility at 
 $Z_a/Z_p = 0.5425$ fixed with the ansatz (\ref{chifit2}). Note
 that the ansatz (\ref{chifit2}) contains no term for analytic 
 corrections.}
  \vskip 0.2cm
  \begin{center}
  \begin{tabular}{|c|c|c|c|}
  \hline
  $L_{min}$ & $\chi^2/$d.o.f. & $\eta$  &  d  \\
  \hline
  12    & 3.92      & 0.03357(15) & 0.9670(4)\phantom{0}  \\
  14    & 2.10      & 0.03431(18) & 0.9688(5)\phantom{0}  \\
  16    & 1.35      & 0.03467(22) & 0.9700(7)\phantom{0}  \\
  20    & 1.30      & 0.03497(28) & 0.9710(9)\phantom{0} \\
  24    & 1.32      & 0.03534(37) & 0.9724(13) \\
  28    & 1.60      & 0.03545(43) & 0.9728(15) \\
\hline
\end{tabular}
\end{center}
\end{table}

Finally we checked for systematic errors due to residual leading order
corrections to scaling at $\lambda=1.1$. For that purpose we fitted 
the magnetic susceptibility at $Z_a/Z_p=0.5425$  
for $\lambda=0.4, 0.8, 1.5$, and $2.5$ with 
$L_{min}=6$ using the ansatz (\ref{chifit1}).  The results  are 
$\eta=0.0493(4)$, $0.0407(6)$, $0.0353(5)$, and $0.0330(5)$ 
for the four values of $\lambda$, respectively. From the results
for $\lambda=0.8$ and $\lambda=1.5$ we get the estimate 
\begin{equation}
 \frac{\Delta \eta_{eff}}{\Delta \lambda} \approx -0.01 \;\; ,
\end{equation}
where 
$\eta_{eff}$ denotes the numerical result for $\eta$ obtained from 
fitting data for lattice sizes $L=6$ up to $L=24$.

From the previous section we know that the difference of $1.1$ and
the values of $\lambda$ where leading order corrections vanish exactly
should be smaller than $0.02$. Therefore the systematic error in our
final estimate of $\eta$ due to residual leading order corrections should
be smaller than $0.01 \times 0.02 = 0.0002$. Note that the lattice
sizes used to obtain our final result range from $L=8$ up to $L=96$.



As final estimate for $\eta$ we take the result from fitting the magnetic 
susceptibility  at $Z_a/Z_p=0.5425$ with the ansatz 
(\ref{chifit1}) and $L_{min}=12$
\begin{equation}
\eta=0.0358(4)[5] \;.
\end{equation}
The estimate of the systematic error is given in the second bracket. It
is obtained from the comparison with the ansatz  (\ref{chifit2})
without analytic corrections
and from the discussion above on $L^{-\omega}$ corrections.

\subsection{The exponent $\nu$}

We compute the exponent $\nu$ from the slope of the Binder cumulant 
and the slope of $Z_a/Z_p$ at $Z_a/Z_p=0.5425$ fixed. 
We fitted our data for $\lambda=1.1$ with the simple power law ansatz
\begin{equation}
\label{simplenu}
\overline{\frac{ \partial U }{\partial  \kappa}} = c \;\; L^{1/\nu} \;\; .
\end{equation}
In table \ref{nu1} we give our fit results for various values of 
the minimal lattice size $L_{min}$ that is included in the fit.
The $\chi^2$/d.o.f. becomes reasonably small for $L_{min}=12$.  Also 
the result for $\nu$ stays stable when $L_{min}$ is further increased.

In table \ref{nu2} we give the analogous results for the slope of 
$Z_a/Z_p$. We see that the $\chi^2$/d.o.f. are much larger than for 
the slope of the Binder cumulant. The value obtained for $\nu$ 
is increasing with increasing $L_{min}$. 
\begin{table}[h]
\caption[blabla8]
{
\label{nu1} \sl
Fits of the derivative eq.  of Binder cumulant $U$ at  $Z_a/Z_p = 0.5425$ 
fixed. The ansatz  (\ref{simplenu}) is used. As results we give the 
exponent of the correlation length $\nu$ and the constant $c$.
}
\vskip 0.2cm
\begin{center}
\begin{tabular}{|c|c|c|c|}
\hline
  $L_{min}$ & $\chi^2/$d.o.f. &  $\nu$  & $2 \; c$ \\
\hline
 \phantom{0}6  &   2.10  &  0.6289(1) & --1.744(1) \\
 \phantom{0}8  &   1.70  &  0.6292(1) & --1.748(2) \\
   10  &   1.76  &  0.6293(2) & --1.750(2) \\
   12  &   1.48  &  0.6296(3) & --1.753(3) \\
   14  &   1.50  &  0.6299(3) & --1.757(4) \\
   16  &   1.76  &  0.6299(4) & --1.756(6) \\
\hline
\end{tabular}
\end{center}
\end{table}

\begin{table}[h]
\caption[blabla8]
{
\label{nu2} \sl
Fits of the derivative of $Z_a/Z_p$  at  $Z_a/Z_p = 0.5425$ fixed.
The ansatz  (\ref{simplenu}) is used. As results we give the
exponent of the correlation length $\nu$ and the constant $c$.
}
\vskip 0.2cm
\begin{center}
\begin{tabular}{|c|c|c|c|}
\hline
 $L_{min}$ & $\chi^2/$d.o.f. &  $\nu$  & $2 \; c$ \\
\hline
\phantom{0}6& 117.0\phantom{00} & 0.6254(1) & --1.1524(4)\phantom{0} \\
\phantom{0}8 &  15.1\phantom{0}  & 0.6273(1)  & --1.1673(5)\phantom{0}\\ 
   10     &   4.9    &  0.6281(1)  & --1.1742(7)\phantom{0} \\
   12     &   2.9   &  0.6286(1)  & --1.1794(12) \\
   14     &   2.2   &  0.6290(2)  & --1.1830(15) \\
   16     &   1.9   &  0.6292(2)  & --1.1858(20) \\
\hline
\end{tabular}
\end{center}
\end{table}

Also for the spin 1 Ising model the authors of ref. \cite{us}
found that the slope 
of the Binder cumulant is scaling much better than the slope of 
$Z_a/Z_p$.

Therefore we take the result obtained from fitting the slope of the 
Binder cumulant with $L_{min}=12 \;\;$
$\nu=0.6296(3)$ as our final result.
In order to check for systematic errors due to residual $L^{-\omega}$ 
corrections we fitted our data for the slope of the Binder 
cumulant at $\lambda=0.4,0.8,1.5$, and $2.5$ with $L_{min}=8$. We obtain 
$\nu=0.6363(3)$, 
$0.6325(5)$,
$0.6271(4)$, and 
 $0.6241(4)$ for the four values of $\lambda$, respectively.
From $\lambda=0.8$ and $1.5$ we obtain 
\begin{equation}
 \frac{\Delta \nu_{eff}}{\Delta \lambda} \approx -0.01 \;\; .
\end{equation}
$\nu_{eff}$ means here the exponent $\nu$ obtained from fitting 
lattices of size $L=8$ up to $L=24$ with the simple power law 
eq. (\ref{simplenu}). 

From the previous section we know that the difference of $1.1$ and 
the values of $\lambda$ where leading order corrections vanish exactly 
should be smaller than $0.02$. Therefore the systematic error in our 
final estimate of $\nu$ due to residual leading order corrections should 
be smaller than $0.01 \times 0.02 = 0.0002$. Also note that the lattice 
sizes used to obtain our final result range from $L=12$ up to $L=96$.
We arrive at the final result
\begin{equation}
\nu=0.6296(3)[4] \;\; ,
\end{equation}
where the second bracket gives an estimate of systematic errors. It is 
obtained from the discussion on residual $L^{-\omega}$ corrections and 
from the comparison of fits of the slope of the Binder cumulant
and of the slope of $Z_a/Z_p$.


\section{Comparison with the literature}
There exist exhaustive compilations of results for critical exponents
in the literature. See for example refs. \cite{guzi,blh}.
In table \ref{literature} we give only most recent results that
reflect the state of the art. 
In ref. \cite{us} the spin-1 model was 
simulated at parameters where leading order corrections vanish (up to 
numerical uncertainties). The approach to compute critical
exponents is very similar to the one 
of the present work. The results for $\nu$ and $\eta$ are consistent 
with the results of this study. The error-bars are of similar size.
In ref. \cite{us} the authors  found indications that the value 
of $\omega$ should be larger than that obtained by field-theoretic methods,
however they were not able to give reliable error 
estimates. Simulating the $\phi^4$ model at 
various values of the coupling constant $\lambda$ allowed us to vary the 
strength of the corrections to scaling. Therefore we 
were able to give a reliable error-estimate for $\omega$.

In ref. \cite{spainparisi} the standard Ising model was simulated. The 
authors also use finite size scaling techniques to compute the exponents.
In the analysis of the data leading order corrections to scaling are 
taken into account. The final result for $\nu$ is perfectly 
consistent with our present result.  The error-bars are a little larger 
than ours. The value for $\eta$ is larger than ours but still compatible 
when the statistical and systematical errors are taken into account.
Their value for the correction exponent $\omega=0.87(9)$ is consistent 
with ours but the error-bar is nine times larger than ours.

Refs. \cite{spainparisi,us} as well as the present work employ finite size 
scaling methods that were pioneered by Binder \cite{binder}. 
It seems to us 
that such an approach is more robust than the so called 
Monte Carlo renormalization  group method  \cite{mcrg}.

 The $\epsilon$-expansion was invented by Wilson and Fisher \cite{WiFi}. 
 Most recent results obtained from the $\epsilon$-expansion and from
 perturbation theory in 3 dimensions are given in ref. \cite{guzi}.
 The results for $\nu$ and $\eta$ obtained from the $\epsilon$-expansion
 are in perfect agreement with our results. However the error-bars are 
 considerably larger than those obtained now from Monte Carlo.
 The result for $\omega$ is smaller than ours but still 
 consistent within error-bars. 
 Parisi \cite{Parisi}
 proposed to perform perturbative expansions directly in 
 three dimensions (3D PT). 
 The  result given in ref.\ \cite{guzi}
 for $\nu$ is consistent with ours.
 The value of $\eta$ is smaller than ours, but still within the quoted 
 error-bars. In both cases our error-bars are considerably smaller.
 The value of the correction exponent 
 $\omega$  obtained from 3D PT \cite{guzi}
 is considerably smaller than our result.
 The results are inconsistent when the error-bars that are quoted are
 taken into account.

 The analysis of high temperature expansions was the first theoretical 
 method that produced reliable, non-mean-field  values for critical
 exponents of three dimensional systems. For a review of early 
 work see ref. \cite{fisher}.
 Still this method gives results that are competitive in accuracy with 
 Monte Carlo and field theoretic methods.
 The authors of ref. \cite{buco} claim to have the most accurate results 
 obtained from  high temperature series expansions of the Ising model 
 on the bcc lattice.  Their result for $\nu$ is larger than ours. The 
 quoted error-bars do not overlap by a small margin. The authors 
 do not quote a value for $\eta$ in their paper. For the convenience of 
 the reader we converted  $\gamma=1.2384(6)$  given for the
 exponent
 of the magnetic susceptibility using the scaling relation 
 $\eta=2-\gamma/\nu$. 
(Converting our numbers we get $\gamma=1.2367(11)$.)
One should note that in this work $\omega$ as well
as $\beta_c$ is taken as external input for the analysis.
 
We also give results obtained from the series expansion of two
parameter models that interpolate between the Gaussian and 
the Ising model, similar to the model considered in the present work.
While the results of Nickel and Rehr \cite{NiRe}
are in perfect agreement with our 
numbers (with larger error-bars) the value obtained by Chen, Fisher and 
Nickel \cite{ChFiNi}  for $\nu$  is clearly larger than ours.
Also the value $\gamma=1.237(2)$  obtained by Nickel and Rehr is 
consistent with ours, while $\gamma=1.2395(4)$ of Chen, Fisher and
Nickel seems too large compared with our result.

The values obtained by Nickel and Rehr as well as by 
Chen, Fisher and Nickel
for the correction to scaling exponent $\omega$ 
are consistent with our result. However the error-bars are considerably 
larger than ours.

Further methods to compute critical exponents 
("exact renormalization group", "coherent anomaly method",  ...) 
are discussed in the literature.
These methods  tend to
give less accurate results and are therefore not discussed here.

\begin{table}[h]
\caption[blabla10]
{
\label{literature} \sl
Recent results for critical exponents obtained with Monte Carlo
simulations (MC), $\epsilon$-expansion, Perturbation-Theory
in three dimensions
(3D,PT) and High temperature series expansions.   When only $\nu$ and 
$\gamma$ are given in the reference we computed $\eta$ with the scaling 
law. These cases are indicated by $^*$.
For a discussion see the text.
}
\vskip 0.2cm
\begin{center}
\begin{tabular}{|c|c|l|l|l|}
\hline
Ref.               & Method &\mc{1}{|c|}{$\nu$} & \mc{1}{|c|}{$\eta$} & 
		    \mc{1}{|c|}{$\omega$} \\
\hline
present work       & MC     &  0.6296(3)(4) & 0.0358(4)(5) &0.845(10)\\
\cite{us}          & MC     &  0.6298(5)    & 0.0366(8)  &         \\
\cite{spainparisi} & MC     &  0.6294(5)(5) & 0.0374(6)(6) & 0.87(9) \\
\cite{guzi}        & 3D,PT    &0.6304(13)& 0.0335(25)  & 0.799(11)  \\
\cite{guzi}        & $\epsilon$,bc &0.6305(25)&0.0365(50)&  0.814(18) \\
\cite{guzi}        & $\epsilon$,free &0.6290(25)&0.0360(50)&0.814(18) \\
\cite{buco}        &  HT      &0.6308(5) &$0.0368(18)^*$&    \\
\cite{ChFiNi}      &  HT      &0.632(1)  &$0.0388(32)^*$&$0.854(80)^*$\\
\cite{NiRe}        &  HT      &0.6300(15)&$0.0365(56)^*$&$0.825(50)^*$\\
\hline
\end{tabular}
\end{center}
\end{table}

Experimental results have been obtained for 
binary mixtures of fluids, vapor-fluid systems and uni-axial 
(anti-)ferromagnetic systems. In general the results agree with the 
predictions yielded by the theoretical methods discussed
above. However the results for the exponents are less accurate than 
the theoretical ones. To give only a few examples: 
The study of the heat capacity of a aniline-cyclohexane mixture
gives $\alpha=0.104(11)$ \cite{ex1}. Using the scaling relation 
$d \nu=2-\alpha$ we get $\nu=0.6320(37)$.  A neutron scattering 
measurement of an antiferromagnetic $\mbox{FeF}_2$ system gave
$\nu=0.64(1)$ and $\gamma=1.25(2)$ \cite{BeYo}.
For a large collection of experimental results see ref. \cite{blh}.

\section{Conclusion and outlook}
In this study we determined with high precision 
the value of the coupling $\lambda$ for which leading corrections
to scaling vanish. This allowed us to obtain results for critical 
exponents with an accuracy better than that of field theoretic methods.

The knowledge of the optimal $\lambda$ can also be used for the 
study of other universal properties of the model, e.g. 
universal amplitude ratios, the effective potential.

The programme to eliminate leading order corrections 
can also be applied to the $\phi^4$ model with more 
than one component of the field.
The case of two components is of particular interest since it is 
supposed to be in the same universality class as superfluid helium 
systems. For this system there exist experimental results for $\nu$ 
\cite{helium}
that are by far more accurate than the existing theoretical predictions.

It would be desirable to extend the programme to subleading corrections.
However these corrections are by far less well understood than  
leading corrections to scaling. Therefore it is unclear how many and what
kind of terms should be included into the action.

However,
even if it is not possible to remove higher order corrections to scaling 
completely
it would be a good check of the reliability of the results to redo 
the study with a different lattice or with more than a next to nearest 
neighbour coupling.

\section{Acknowledgements}
We thank K. Pinn and T. T\"or\"ok for discussions and comments on the 
manuscript.

\end{document}